# A New Energy Ordering of Gas Phase Glycine and the Dipole Moment via Plane-wave Density Functional Theory Calculations

**Byeong June Min**

*Department of Physics, Daegu University, Kyungsan 712-714, Korea*

The abundance of glycine (Gly), the simplest amino acid, in meteorites leads us to the next question about its extraterrestrial origin. However, astronomers have not yet found glycine signature in interstellar medium. Laboratory microwave spectroscopy experiments report the most stable Gly conformer has a dipole moment of 4.5 – 5.45 Debye. Theoretical calculations, so far performed only with Gaussian basis functions, has predicted a dipole moment of about 1 Debye. This discrepancy has baffled astronomers. We study the energetics of glycine and its isomers and conformers via plane-wave density functional theory calculations. The geometric structures of the isomers and their conformers are identified, along with their relative stability and their dipole moment. In the case of glycine, we obtain the most stable conformer with a dipole moment of 5.76 Debye, close to the microwave spectroscopy experiments. If the plane wave energy cutoff is reduced to a lower value (~400 eV) on purpose, the energy ordering reverses to the case with Gaussian basis calculations.

Email: bjmin@daegu.ac.kr

Fax: +82-53-850-6439, Tel: +82-53-850-6436

# I. INTRODUCTION

Discovery of amino acids in meteorites [1] strongly suggest extraterrestrial origin of amino acids. Glycine (Gly, $NH_2CH_2COOH$) as the simplest amino acid would be one of the earliest amino acids generated in space. Astronomers have been trying to detect the presence of glycine in interstellar medium [2, 3]. The search, so far, has not been successful [3], but the detection of acetic acid ($CH_3COOH$) in interstellar clouds [4, 5] strengthens the likelihood of its presence. Relative abundance of Gly isomers in interstellar space also needs to be considered. Thus, the energetics of Gly and its isomers, N-methylcarbamic acid $CH_3NHCOOH$ and methyl carbamate $CH_3OC(NH_2)O$, and their conformers is of great importance.

It may seem plausible that Gly, being the smallest amino acid and one of the basic building blocks of proteins that constitute life around us, would have the lowest energy among its isomers. However, that does not turn out to be true [3, 6]. N-methylcarbamic acid (NMCA) is the most stable isomer, methyl carbamate (MC) the next stable isomer, and Gly is the least stable isomer among these three. This fact must add to the difficulty of generating life.

Gly itself has long been the subject of many theoretical studies [6 - 11]. However, there has been no plane-wave based studies. We study neutral Gly, NMCA, and MC, and their conformers via plane-wave density functional theory calculations. The geometric structures, the energy ordering, and the dipole moment are determined and compared with existing calculations and experiments. The harmonic vibration frequencies were calculated and the zero-point energy is taken into account when the relative stability is estimated.

Earlier theoretical studies were performed using Gaussian basis functions to represent the electronic wavefunctions. In an effort to refine the results, researchers have been concentrating on improving the results by introducing perturbative corrections due to electronic correlations, assuming that the zeroth order calculations are accurate enough. However, we find that the energy ordering of the Gly conformers changes at a smaller plane-wave energy cutoff. Such reversal of energy ordering was not observed for NMCA and MC. This may have to do with Gly not being the most stable isomer, but only the third.

The dipole moment of the most stable conformer of Gly has far implications in its interstellar search. So far, quantum chemistry calculations predicted a dipole moment of about 1 Debye. This result goes against microwave spectroscopy measurement of 4.5 – 5.45 Debye [12, 13]. In our calculations, the most stable Gly conformer has a dipole moment of about 5.76 Debye, resolving the contradiction.

## II. CALCULATION

The calculations were performed using the ABINIT package [14] with periodic boundary condition and gamma point sampling. Norm-conserving pseudopotentials were used [15]. Plane wave energy cutoff of 1200 eV is used. The cubic box dimension was chosen as 13.2 Å. The total-energy converges within 1 meV. Exchange correlation energy was described by the Perdew-Burke-Ernzerhof parameterization within the generalized gradient approximation (PBE-GGA) [16]. Semi-empirical treatment of the van der Waals interaction is used [17].

Self-consistency cycles were repeated until the difference of the total energy becomes smaller than $2.7 \times 10^{-6}$ eV, twice in a row. The system was relaxed until the average force on the atoms becomes smaller than $2.3 \times 10^{-3}$ eV/A by Broyden-Fletcher-Goldfarb-Shanno (BFGS) minimization scheme [18].

Dipole moment $\vec{P}$ is calculated from the differential electron density $\rho_{\text{diff}}(\vec{r}) = \rho(\vec{r}) - \sum \rho_{\text{atom}}(\vec{r})$, as $\vec{P} = -e \int \vec{r} \rho_{\text{diff}}(\vec{r}) d^3\vec{r}$ [19] rather than $\vec{P} = -e \int \vec{r} \rho(\vec{r}) d^3\vec{r} + e \sum Z_i \vec{R}_i$. We observed that these two expressions yield almost identical results, but chose to use the former because this would enable us to visualize the polarization distribution in the molecule.

## III. RESULTS AND DISCUSSION

1. **Relative Energies of the Most Stable Conformers of NMCA, MC, and Gly**

In agreement with previous theoretical calculations [3, 6, 10], we also find that NMCA, not Gly, is the most stable isomer. Relative to the NMCA, MC is 0.136 eV higher, and Gly 0.376 eV higher. When the zero-point energy is considered, these become 0.112 eV and 0.385 eV, respectively. Previous studies by Kayi *et al.* [6] and Lattelais *et al.* [3, 10] performed both Becke, three-parameter, Lee-Yang-Parr hybrid functional (B3LYP) and perturbated triple excitation coupled cluster [CCSD(T)] calculations using various sets of Gaussian basis functions. Their CCSD(T) results show opposite tendency compared to the B3LYP hybrid functional results: Kayi *et al.* report Gly energy as 0.458 eV (B3LYP) and 0.355 eV [CCSD(T)], but Lattelais *et al.* report 0.385 eV (B3LYP) and 0.464 eV [CCSD(T)]. Our Gly energy is comparable to these results, but the MC energy 0.112 eV is smaller almost by half compared to 0.214 eV (B3LYP) and 0.206 eV [CCSD(T)] by Lattelais *et al*.

2. **Conformers of NMCA**

We find that the $CH_3$ group rotation in NMCA involves negligible amount of energy. Then, we are left with four low-energy conformers of NMCA (Figure 1). All four conformers have planar structures except the internal rotation of the $CH_3$ group. Their relative energies are summarized with respect to the most stable conformer (NMCA-1) in Table 1. Zero-point energies (ZPE) are included.

The relative energy of NMCA-2 is 0.055 eV from our calculations, compared to 0.058 eV [CCSD(T)] and 0.061 eV (B3LYP) from Kayi *et al*. and 0.065 eV (B3LYP) from Lattelais *et al*. The correction from the electron correlation as described in CCSD(T) is quite small. The typical difference in the relative energy is about 6 %. However, the energy difference grows very fast for NMCA-3 and NMCA-4, to about 9% and 18 %, respectively.

We also calculated the dipole moment from the differential electron density. Dipole moment of the NMCA-1 is 3.28 Debye, compared to 2.465 Debye from MP2 calculations by Kayi *et al*. [6] and 2.4 Debye from B3LYP calculations by Lattelais *et al*. [3]

### 3. Conformers of MC

We have two conformers for methyl carbamate. Again, all the atoms except the H's in the $CH_3$ group are coplanar (Figure 2). The relative energy difference is 0.252 eV, compared to 0.347 eV from B3LYP calculations by Lattelais *et al*. This corresponds to almost 30 % difference. The zero-point energy, in this case, is practically the same for the two conformers, contributing only 1 meV difference.

The most stable MC conformer has a dipole moment of 3.35 Debye, compared to 7.01 Debye of the higher energy conformer. Lattelais *et al*. report 2.4 Debye and 5.1 Debye, respectively.

### 4. Conformers of Gly

Eight lowest energy conformers of Gly are shown in Figure 3 in the increasing energy order. All conformers have coplanar arrangement of C, O, N atoms, except Gly-5 and Gly-8, in which the COO group is rotated slightly out of the plane formed by $N-C_1-C_2$. Our results coincide with Csaszar's self-consistent field (SCF) level results [8] in that only Gly-5 and Gly-8 are non-planar conformers.

However, our results are at variance with the previous DFT calculations performed at the SCF level with Gaussian basis functions [6 - 11], regardless of the specific choice of Gaussian basis functions. All previous theoretical calculations based on Gaussian functions unanimously predict Gly-2 or Gly-4 as the most stable conformer, not Gly-1 of our calculations. The origin of this energy ordering reversal is traced back to the insufficient plane-wave energy cutoff. As we repeat the calculation using smaller plane-wave energy cutoff, we find the Gly-2 becomes the most stable conformer below the plane-wave energy cutoff of 400 eV. Thus, the potential energy surface of Gly undergoes a succinct transformation as the plane-wave energy cutoff changes and the relative energies of the conformers get shuffled. The convergence of the energy difference $\Delta E = E(Gly-2) - E(Gly-1)$ and the dipole moment $\mu_1$ of Gly-1 and $\mu_2$ of Gly-2 against the plane-wave energy cutoff is shown in Figure 4. The differential electron density profiles $\rho_{diff}(\vec{r}) = \rho(\vec{r}) - \sum \rho_{atom}(\vec{r})$ for Gly-1 and Gly-2 are plotted in Figure 5 and Figure 6.

Our prediction of the most stable conformer agrees well with the microwave spectroscopy experiments by R. D. Brown et *al.* [12], in which the authors suggested the current Gly-1 structure as the most stable conformer in contradiction to the theoretical calculations available at that time. R. D. Brown et *al.* also report the dipole moment of the most stable conformer to be 4.5 – 4.6 Debye. More recent experiment by F. J. Lovas *et al.* report 5.45 Debye [13]. The most stable conformer predicted by Csaszar is Gly-2 with a dipole moment of 1.29 Debye (correlated level), or Gly-4 with 1.99 Debye (SCF level). In our calculations, the dipole moment of Gly-1 and Gly-2 are 5.76 Debye and 1.19 Debye, respectively. Our results are in excellent accord with the microwave experiment and resolve the discrepancy between the theory and the experiment. Comparison of the relative energies of Gly conformers and the dipole moment is summarized in Table 2. The geometries and dipole moment of Gly conformers are summarized and compared with other calculations [11] in Table 3.

**IV. CONCLUSION**

We studied the energetics of Gly isomers and conformers via plane-wave density functional theory calculations. Gly is higher in energy than its isomers NMCA and MC. The most stable conformer of Gly has a dipole moment of 5.76 Debye, consistent with microwave spectroscopy experiments. This resolves the dilemma caused by previous theoretical prediction of the most stable conformer with a dipole moment of about 1 Debye.

ACKNOWLEDGMENTS

This research was supported in part by the Daegu University Research Funds.

REFERENCES

[1] See, for example, a review by A. S. Burton, J. C. Stern, J. E. Elsila, D. P. Glavin, and J. P. Dworkin, Chem. Soc. Rev. **41**, 5459 (2012).

[2] J. M. Hollis, J. A. Pedelty, L. E. Snyder, P. R. Jewell, F. J. Lovas, P. Palmer, and S.-Y. Liu, Astrophys. J. **588**, 353 (2003).

[3] M. Lattelais, F. Pauzat, J. Pilme, Y. Ellinger, and C. Ceccarelli, Astron. Astrophys. **532**, A39 (2011).

[4] D. M. Mehringer, L. E. Snyder, Y. Miao, and F. J. Lovas, Astrophys. J. **480**, L71 (1997).

[5] A. J. Remijan, L. E. Snyder, S. -Y. Liu, D. M. Mehringer, and Y. -J. Kuan, Astrophys. J. **576**, 264 (2002).

[6] H. Kayi, R. I. Kaiser, and J. D. Head, Phys. Chem. Chem. Phys. **13**, 15774 (2011).

[7] J. H. Jensen and M. S. Gordon, J. Am. Chem. Soc. **113**, 7917 (1991).

[8] A. G. Csaszar, J. Am. Chem. Soc. **114**, 9568 (1992).

[9] C. H. Hu, M. Shen, and H. F. Schaffer III, J. Am. Chem. Soc. **115**, 2923 (1993).

[10] M. Lattelais, Y. Ellinger, and B. Zanda, Int. J. Astrobiol. **6**, 37 (2007).

[11] H. L. Sellers and L. Schäfer, J. Am. Chem. Soc. **100**, 7728 (1978).

[12] R. D. Brown, P. D. Godfrey, J. W. V. Storey, and M.-P. Bassez, J. Chem. Soc. Chem. Commun. 547 (1978).

[13] F. J. Lovas, Y. Kawashima, J.-U. Grabow, R. D. Suenram, G. T. Fraser, and E. Hirota, Astrophys. J. **455**, L201 (1995).

[14] X. Gonze, B. Amadon, P. M. Anglade, J. M. Beuken, F. Bottin, *et al*, Computer Phys. Commun. **180**, 2582 (2009). The ABINIT code is a common project of the Université Catholique de Louvain, Corning Incorporated, and other contributors (URL http://www.abinit.org, accessed January 2014).

[15] D. R. Hamann, Phys. Rev. B 88, 085117 (2013).

[16] J. P. Perdew, K. Burke, and M. Ernzerhof, Phys. Rev. Lett. **77**, 3865 (1996).

[17] S. Grimme, J. Comput. Chem. **27**, 1787 (2006).

[18] W. H. Press, B. P. Flannery, S. A. Teukolsky, and W. T. Vetterling, *Numerical Recipes* (Cambridge University Press, New York, 1986).

[19] B. J. Min, arXiv: 1708.03834 [physics.chem-ph].

Table 1. Zero-point energy corrected relative energies (eV) and dipole moment (Debye) of N-methylcarbamic acid conformers, compared with other theoretical calculations.

| | Relative energy (present) | Dipole moment (present) | Relative energy Kayi CCSD(T) [6] | Relative energy Kayi B3LYP [6] | Dipole moment Kayi MP2 [6] | Relative energy Lattelais B3LYP [3] | Dipole moment Lattelais B3LYP [3] |
|---|---|---|---|---|---|---|---|
| NMCA-1 | 0 | 3.28 | 0 | 0 | 2.465 | 0 | 2.4 |
| NMCA-2 | 0.055 | 3.65 | 0.058 | 0.061 | 2.826 | 0.065 | 2.6 |
| NMCA-3 | 0.262 | 6.77 | 0.285 | 0.294 | 4.824 | 0.343 | 4.9 |
| NMCA-4 | 0.278 | 7.39 | 0.341 | 0.335 | 5.556 | 0.386 | 5.4 |

Table 2. Comparison of the relative energy (eV) and dipole moment (Debye) of Gly conformers with previous density functional theory calculations.

|       | Relative energy (present calculation) | Dipole moment (present calculation) | Relative energy (SCF, Csaszar) | Dipole moment (SCF, Csaszar) | Relative energy (Lattelais *et al.*) | Dipole moment (Lattelais *et al.*) |
|-------|---------------------------------------|-------------------------------------|--------------------------------|------------------------------|--------------------------------------|------------------------------------|
| Gly-1 | 0                                     | 5.76                                | 0.152                          | 6.30                         | 0.035                                | 5.6                                |
| Gly-2 | 0.036                                 | 1.19                                | 0                              | 1.29                         | 0                                    | 1.2                                |
| Gly-3 | 0.086                                 | 2.01                                | 0.217                          | 3.16                         | 0.078                                | 2.0                                |
| Gly-4 | 0.105                                 | 1.90                                | 0.088                          | 1.99                         | 0.069                                | 1.9                                |
| Gly-5 | 0.157                                 | 2.37                                | 0.244                          | 2.85                         | 0.143                                | 2.4                                |
| Gly-6 | 0.192                                 | 4.39                                | 0.371                          | 4.68                         | 0.269                                | 4.3                                |
| Gly-7 | 0.212                                 | 2.96                                | 0.286                          | 3.53                         | 0.239                                | 2.9                                |
| Gly-8 | 0.250                                 | 3.99                                | 0.364                          | 4.92                         | 0.317                                | 4.0                                |

Table 3. Comparison of the geometry (Å and degrees), relative energy $E_{rel}$ (eV) and dipole moment $\mu$ (Debye) of Gly conformers with previous density functional theory calculations by H. L. Sellers and L. Schäfer [11]. (*) is the lowest energy conformer in each calculations.

|  | Gly-1 present(*) | Gly-1 Sellers *et al.* | Gly-2 present | Gly-2 Sellers *et al.*(*) |
|---|---|---|---|---|
| r(N-H) | 1.016 | 1.000 | 1.019 | 1.001 |
| r(N-C) | 1.472 | 1.474 | 1.450 | 1.457 |
| r(C-H) | 1.098 | 1.081 | 1.101 | 1.081 |
| r(C-C) | 1.541 | 1.535 | 1.526 | 1.514 |
| r(C=O) | 1.210 | 1.202 | 1.212 | 1.203 |
| r(C-O) | 1.345 | 1.345 | 1.363 | 1.364 |
| r(O-H) | 1.000 | 0.975 | 0.976 | 0.966 |
| θ(NCC) | 111.06 | 110.19 | 115.53 | 113.28 |
| θ(CC=O) | 122.99 | 122.32 | 125.54 | 126.41 |
| θ(CC-O) | 113.00 | 113.82 | 112.27 | 110.62 |
| θ(CO-H) | 103.13 | 108.44 | 106.55 | 112.28 |
| θ(CNH) | 112.65 | 114.49 | 109.90 | 113.27 |
| θ(CCH) | 107.20 | 107.67 | 107.50 | 107.87 |
| θ(HNH) | 108.47 | 111.36 | 105.87 | 110.29 |
| θ(HCH) | 106.75 | 107.37 | 105.48 | 107.04 |
| θ(NCH) | 111.90 | 111.87 | 110.06 | 110.27 |
| τ(NCC=O) | 179.09 | 180.0 | 0.83 | 0.0 |
| τ(NCCO) | 1.01 | 0.0 | 179.26 | 180.0 |
| τ(CCOH) | 0.31 | 0.0 | 179.89 | 180.0 |
| τ(CCNH) | 120.01 | 114.83 | 57.37 | 63.29 |
| τ(O-CCH) | 121.51 | 122.25 | 57.44 | 57.65 |
| $E_{rel}$ | 0.0 | 0.095 | 0.036 | 0.0 |
| μ | 5.76 | 6.54 | 1.19 | 1.10 |

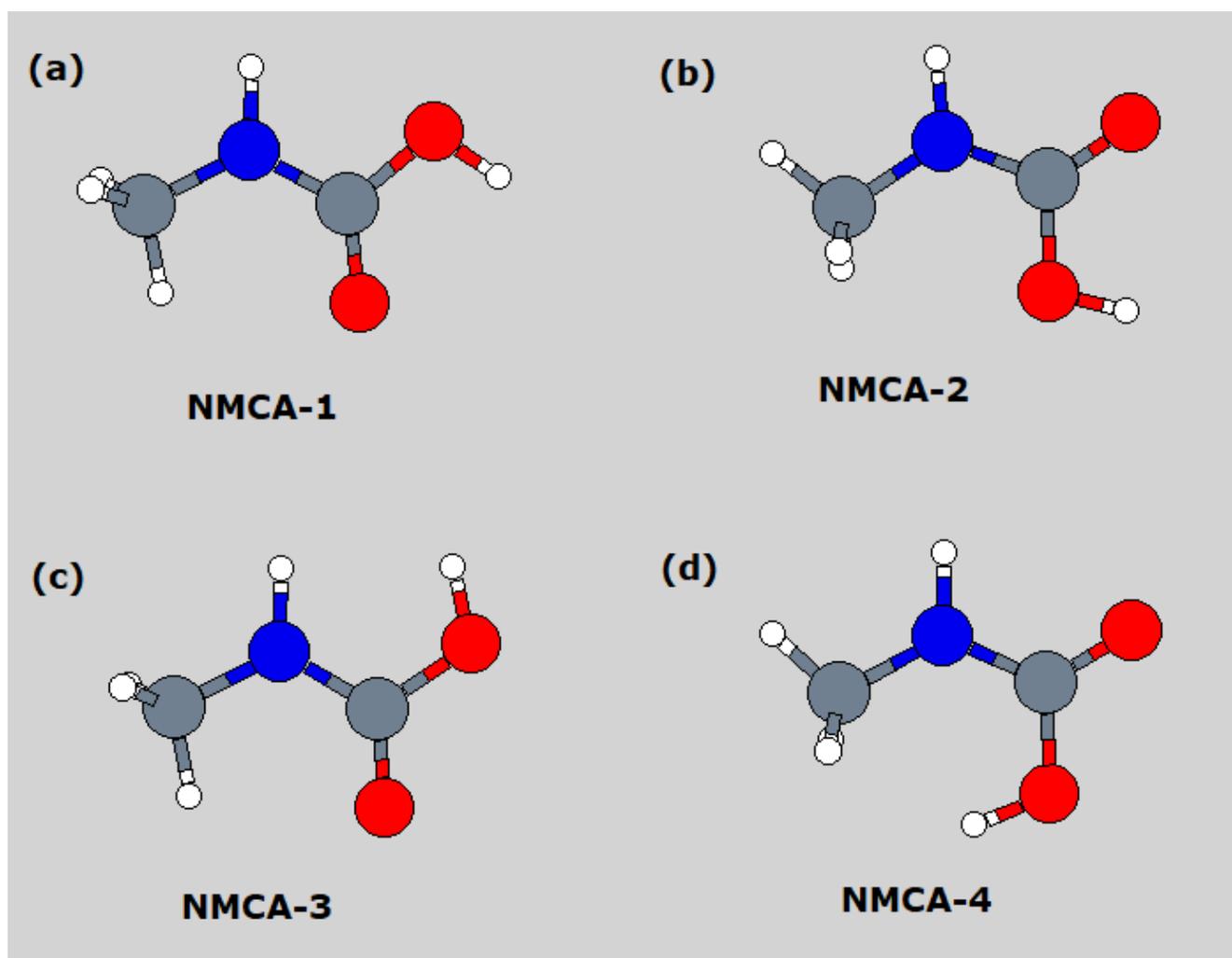

Fig. 1.

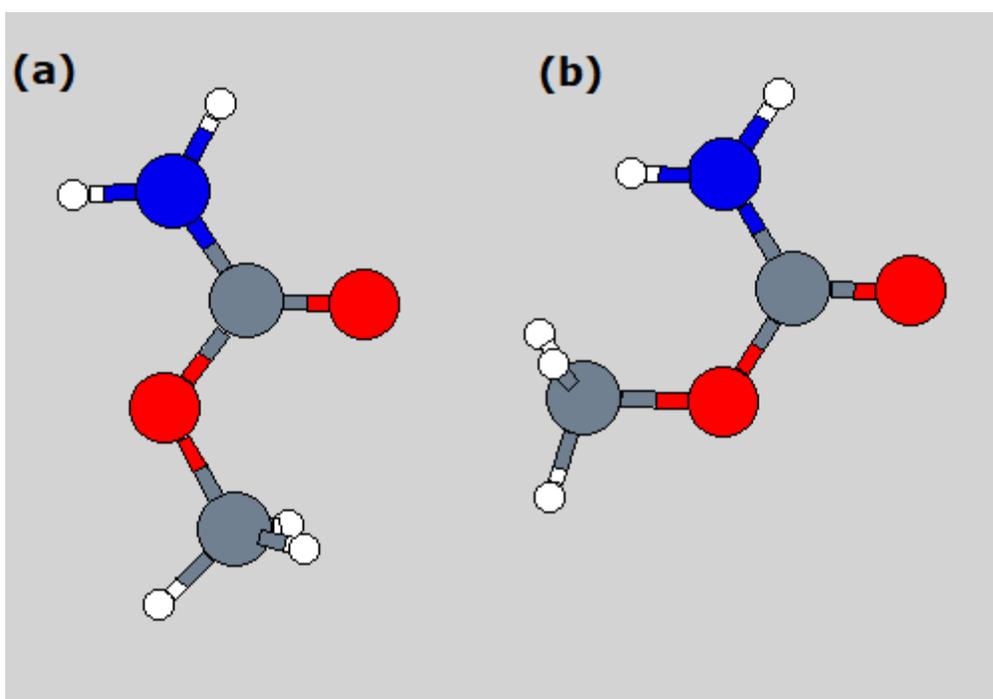

Fig. 2

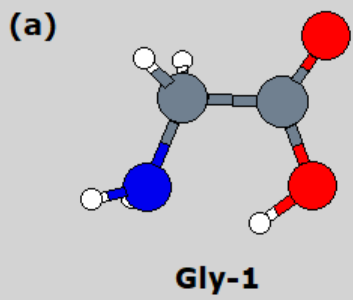
**Gly-1**

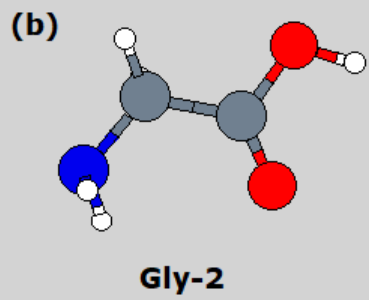
**Gly-2**

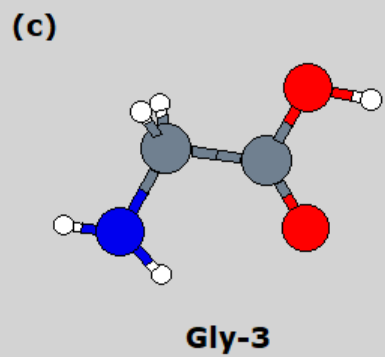
**Gly-3**

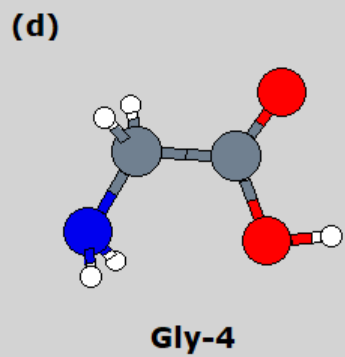
**Gly-4**

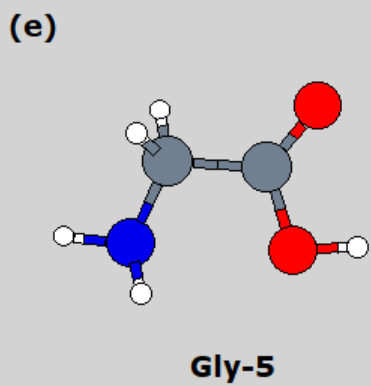
**Gly-5**

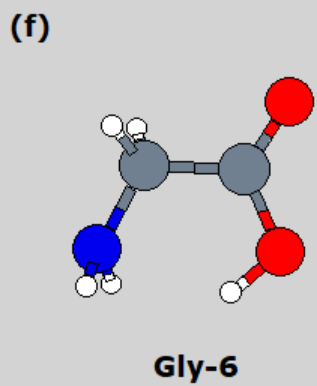
**Gly-6**

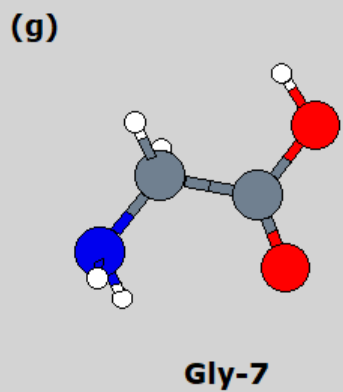
**Gly-7**

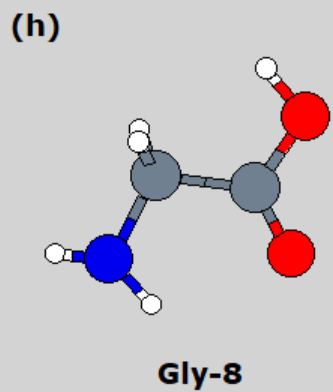
**Gly-8**

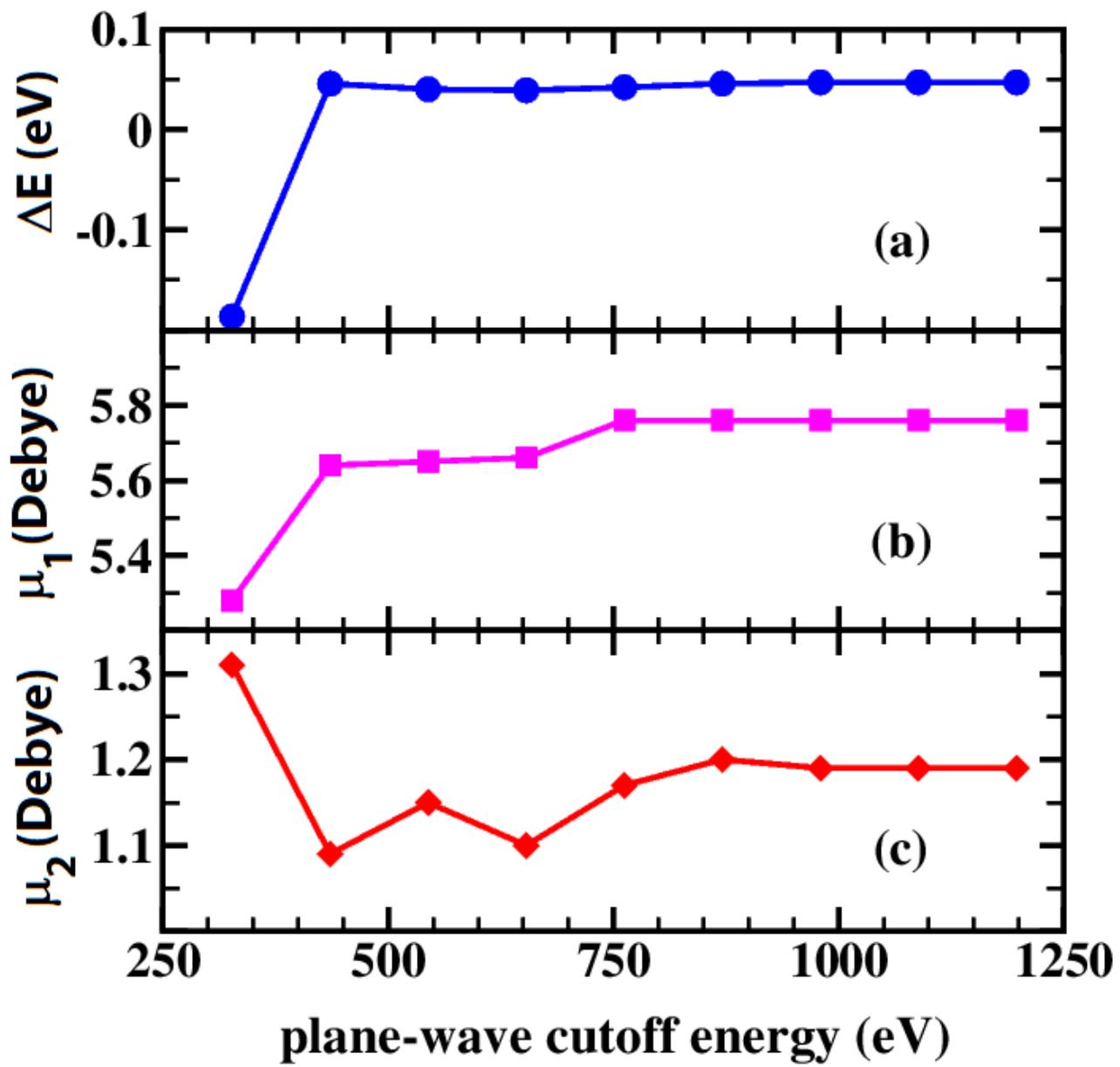

Figure 4

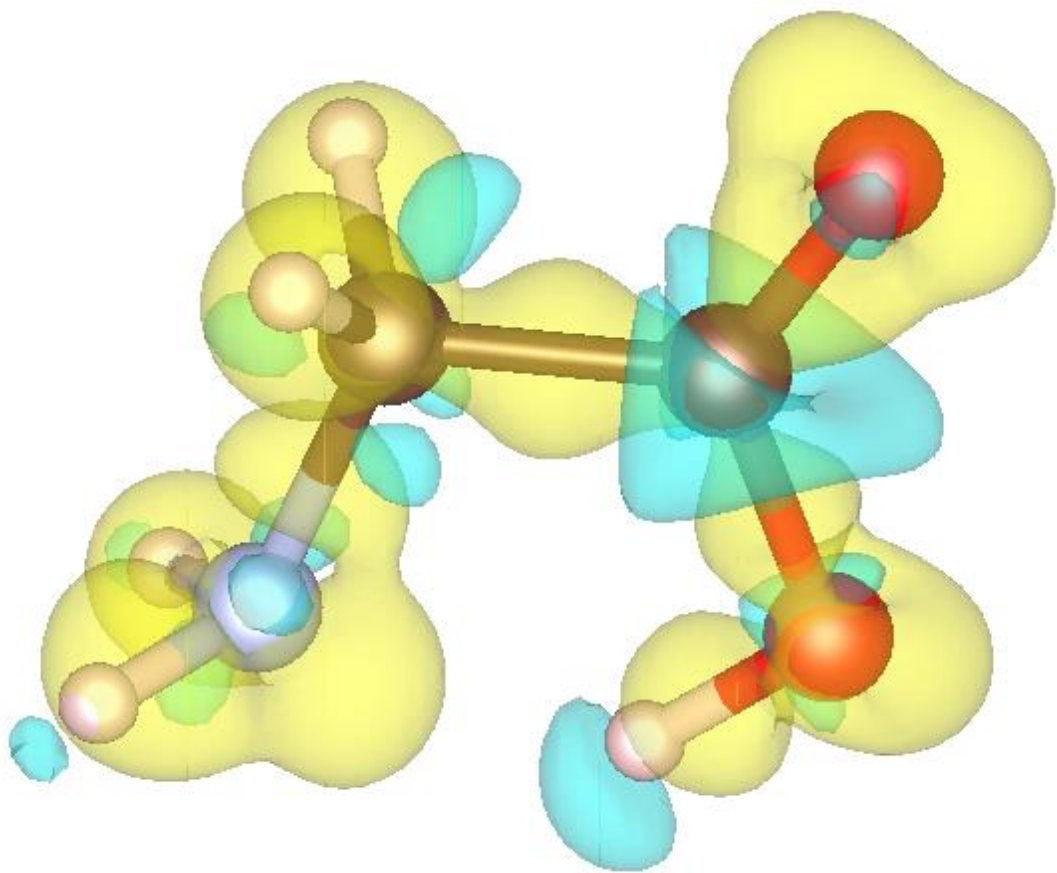

Figure 5

Figure 6

Figure Captions.

Fig. 1. Four conformers of N-methylcarbamic acid in the increasing energy order.

Fig. 2. Two conformers of methyl carbamate in the increasing energy order.

Fig. 3. Eight lowest energy conformers of Gly in the increasing energy order.

Fig. 4. The convergence of the energy difference $\Delta E = E(\text{Gly}-2) - E(\text{Gly}-1)$ and the dipole moment $\mu_1$ of Gly-1 and $\mu_1$ of Gly-2 against the plane-wave energy cutoff.

Fig. 5. The differential electron density profile for Gly-1.

Fig. 6. The differential electron density profile for Gly-2.